# Ultrafast Dynamic Control of Spin and Charge Density Oscillations in a GaAs Quantum Well

J. M. Bao,[1] L. N. Pfeiffer,[2] K. W. West,[2] and R. Merlin[1]

[1]*Focus Center and Department of Physics, The University of Michigan,*

*Ann Arbor, Michigan, 48109-1120*

[2]*Bell Laboratories, Lucent Technologies, Murray Hill, New Jersey, 07974*

We use subpicosecond laser pulses to generate and monitor in real time collective oscillations of electrons in a modulation-doped GaAs quantum well. The observed frequencies match those of intersubband spin- and charge-density excitations. Light couples to coherent density fluctuations through resonant stimulated Raman scattering. Because the spin- and charge-related modes obey different selection rules and resonant behavior, the amplitudes of the corresponding oscillations can be independently controlled by using shaped pulses of the proper polarization.

PACS number: 72.25.Fe, 73.21.Fg, 78.47.+p, 42.65.Dr

Spin currents are the most common source of magnetism and, since spin cannot be described in classical terms, it is apparent that the majority of magnetic phenomena are ultimately a manifestation of quantum behavior. While this fact has been known for a very long time, it is only recently that methods to generate spin-polarized currents have attracted much attention driven mainly by the possibility that novel quantum effects and devices may be uncovered [1]. Next to electrical methods, the generation of magnetic currents by optical injection has now become a very active area of research. Here III-V semiconductor quantum wells (QW's), particularly those belonging to the AlGaAs/GaAs system, play an important role owing to the spin-polarized nature of their valence band [2]. We note that these heterostructures have been used for many years to produce, by means of photoexcitation, incoherent spin-polarized electron sources for applications in nuclear and high energy physics [3]. More recently, injection a pure spin current has been achieved in a GaAs-QW through interference of one- and two-photon absorption processes [4].

In this work, we use ultrafast light pulses to induce coherent density oscillations associated, separately, with the spin and charge degrees of freedom of a quasi-two-dimensional electron gas (2DEG) contained in a single GaAs QW. Studies of the ultrafast dynamics of low-lying levels of a QW have been previously reported [5-14]. This includes work on doped [5-8] and photoexcited QW's as well as on Bloch oscillations [9-11]. Our results distinguish themselves from these studies in that we are able to differentiate collective (many-particle) from single-particle behavior and observe many-electron dynamics in real time (for recent theoretical work on intersubband excitations, see [15,16]). The method we use to generate and control the spin and charge oscillations is stimulated Raman scattering (RS) by intersubband excitations. Given that *spontaneous* RS is one of the main tools for probing 2DEG properties and, in particular, the quantum Hall effects [17-18], our results hold promise for elucidating the coherent dynamics of these and other 2DEG phenomena.

Our sample, grown by molecular beam epitaxy on a (001) GaAs substrate, is a 400-Å one-sided modulation-doped GaAs single QW sandwiched between $Al_{0.3}Ga_{0.7}As$ barriers; see Fig. 1(a). The 2DEG originates from electrons initially bound to those Si donors in the barriers which are closest to and migrate to the QW. To reduce ionized impurity scattering and thereby enhance the mobility, these donor atoms are separated from the 2DEG by a $10^3$-Å-thick undoped spacer [19]. A schematic energy level diagram is shown in Fig. 1(b). The excitations pertinent to our work are intersubband transitions associated with the lowest-lying states of the QW. We used transport measurements at 4.2 K to determine the sample mobility $\mu \approx 2.9 \times 10^6$ cm$^2$/Vs and the 2DEG areal density $\sigma_0 \approx 1.9 \times 10^{11}$ cm$^{-2}$ for which the corresponding Fermi energy is $E_F \approx 7$ meV. The latter value is consistent with the width of the main photoluminescence (PL) feature in Fig. 1(c) [20]. From the PL data, we also get ~ 1.512 eV for the renormalized QW bandgap [21]. Fig. 1(d) shows schematically the long-wavelength limit of the excitation spectrum involving the two lowest-lying subbands. The wavevector $\mathbf{q}$ is perpendicular to the growth axis [001]. The spectrum consists of the single-particle (SP$_{01}$) continuum delimited by $E_{01} \pm \hbar q k_F/m$ and, at small wavevectors, the collective spin-density (SD$_{01}$) and charge-density (CD$_{01}$) resonances [22]. Here, $k_F$ is the Fermi wavevector, $m$ is the electron effective mass and $E_{01} = (E_1 - E_0)$. The dominant direct Coulomb repulsion shifts the plasmon-like charge-density mode to higher energies whereas exchange-correlation effects push the spin-density resonance below the continuum [22]. As shown in Fig. 1(e), Raman data of our QW at $\mathbf{q} = 0$ (vertical transitions) reveal the expected three distinct features at 2.60 (SD$_{01}$), 2.85 (SP$_{01}$) and 3.31 (CD$_{01}$) THz. These results are typical of high-mobility samples [23-25]. Here, $z$ denotes the axis normal to the layers and $x'$ ($y'$) is along the [110] ([1$\bar{1}$0]) direction [26]. We notice that the charge (spin) related peak appears in polarized (depolarized) spectra, i. e., when the incident and scattered polarizations are parallel (orthogonal) to each other. This reflects the fact that the charge (spin) density mode transforms like the symmetric $A_1$ (antisymmetric $A_2$) representation of the $D_{2d}$ point group of the QW [23]. In passing, we note that the 2DEG density can be gained from the measured $SD_{01}$, $SP_{01}$ and $CD_{01}$ frequencies [24-25]. Consistent with transport studies, the RS results give $\sigma_0 = (1.6 \pm 0.2) \times 10^{11}$ cm$^{-2}$ .

Time domain pump-probe experiments were performed at $\sim$ 7K in the reflection geometry using a mode-locked Ti-sapphire laser which provided $\sim$ 50-65 fs pulses at a repetition rate of 82 MHz. The laser beams penetrated the crystal along the $z$–axis (hence, $\mathbf{q} = 0$) and were focused onto a 300-$\mu$m-diameter spot. The average power of the pump beam was 2.5-20 mW (the energy density per pulse was $U = 2$-$10 \times 10^{-8}$ J/cm$^2$), and 2.5 mW for the probe. We measured the differential reflection, defined as the pump-induced change in the reflected probe intensity, as a function of the time delay between the two pulses. The pump beam was either circularly or linearly polarized, along $x'$, while the incident probe beam was linearly polarized, along $y'$. By measuring separately (*i*) the rotation of the polarization angle [27] and (*ii*) the intensity of the probe after reflecting off the sample, we were able to determine the orientation of the scattered beam polarization. To enhance the signal due to the 2DEG, we tuned the central energy of the pulses to a range where the spontaneous $CD_{01}$ Raman cross section exhibits a pronounced enhancement; see inset of Fig. 2. The positions of the maxima, at $\sim$ 1.542 eV and $\sim$ 1.552 eV, are consistent with those of the incoming and outgoing resonances [23] with heavy-hole excitons associated with the third lowest QW state of energy $E_2$.

The pump-probe data, reproduced in Fig. 2, show well-resolved oscillations. After removal of the slow-decaying electronic background, we used linear prediction methods [28] to determine the number of oscillators and their parameters. This procedure gives three modes and fits such as those of Fig. 2 which reproduce quite accurately the experimental traces. The frequencies of the two lowest modes agree extremely well with those of $SD_{01}$ and $CD_{01}$ from the Raman spectra in Fig. 1(e) and, on this basis, we assign them to coherent spin- and charge-density oscillations. The weaker single-particle peak was only vaguely distinguished in the time-domain data. Our experiments show that the $SD_{01}$ mode can only be excited if the pump beam is circularly polarized (Fig. 2, bottom trace) whereas the $CD_{01}$ amplitude is largest for linearly-polarized pulses. This selectivity, as well as the fact that the intersubband oscillation amplitudes depend strongly on the central energy of the pulses opens the road for coherent control studies. $SD_{01}$ and $CD_{01}$ behave also quite differently vis-à-vis the probe detection scheme. While the $CD_{01}$ contribution dominates the modulated intensity (Fig. 2, top trace), $SD_{01}$ leads mainly to a rotation of the probe polarization. Hence, the scattered probe pulses are predominantly perpendicular (parallel) to the incident beam for spin (charge) oscillations. The appearance of the dominant charge-density modes in the bottom trace of Fig. 2 and the depolarized Raman spectrum of Fig. 1(e) is attributed to a polarization leakage. The remaining feature at 3.97 THz, labeled $CD_{12}$, is ascribed to charge-density transitions of *photoexcited* electrons involving the states of energies $E_1$ and $E_2$. This assignment is supported by the calculated QW level spacing, the fact that it exhibits the same selection rules as $CD_{01}$, and by the results depicted in Fig. 3 which show that the intensity of $CD_{12}$ increases with increasing power. Results similar to those of Fig. 3(b) have been reported early in the RS literature [29]. An example of coherent control methods is shown in Fig. 4. Here, the experimental parameters are the same as for the top trace of Fig. 2, but we use two pump pulses of equal intensity to control the amplitudes of the charge-density oscillations. The top trace illustrates destructive interference for $CD_{12}$ in that the two pulses are separated in time by one and one half the $CD_{12}$ period. The fact that the motion does not come to a complete stop is attributed to the increase in the photoexcited electron density caused by the second pulse. In the bottom trace, the two pulses are separated by twice the $CD_{12}$ period, i. e., the interference is constructive and, accordingly, the $CD_{12}$ amplitude is larger. Since the $CD_{01}$ period is $\sim 0.3$ ps, the second pulse has a much weaker effect on this mode.

The correlation we find between time-domain and spontaneous RS results, particularly in regard to the positions of the peaks, the resonant behavior and the selection rules (see below), strongly indicates that stimulated RS is the mechanism responsible for the coherent 2DEG oscillations. Following work on impulsive stimulated RS by phonons [30-31], the coherent interaction between the electromagnetic field and 2DEG density fluctuations is described, with minor modifications, by the same effective Hamiltonian which accounts for spontaneous intersubband RS [32]. Phenomenologically, we write the coupling energy as

$$H_S = \sum_{jl} \int E_j^*(\omega_1) \gamma_{jl}(\omega_1, \omega_2) E_l(\omega_2) d\omega_1 d\omega_2 + \text{c.c.} \qquad (1)$$

where $\mathbf{E}(\omega)$ is the Fourier transform of the (pump or probe) electric field $\mathbf{E}(t)$ and

$$\gamma_{jl} = C_{jl}(\hat{\sigma}_{\mathbf{q}\uparrow} + \hat{\sigma}_{\mathbf{q}\downarrow}) + i S_{jl}(\hat{\sigma}_{\mathbf{q}\uparrow} - \hat{\sigma}_{\mathbf{q}\downarrow}) \ . \qquad (2)$$

Here $\hat{\sigma}_{\mathbf{q}s} = \sum_{\mathbf{k},mn} c_{\mathbf{k}s,n}^+ c_{(\mathbf{k}+\mathbf{q})s,m}$ are density fluctuation operators, $c_{\mathbf{k}s,n}^+$ ($c_{\mathbf{k}s,n}$) is the electron creation (annihilation) operator for the state $|\mathbf{k}s, n\rangle$ of wavevector $\mathbf{k}$, spin component $s$ and subband index $n$, and $C_{jl}$ ($S_{jl}$) is the Raman tensor describing the coupling to charge (spin) density fluctuations (for notation clarity, we ignore the dependence of the coupling constants on the subband index).

Note that $H_S$ commutes with the total spin of the 2DEG, and that the signs of the spin-up and spin-down operators are the same for $C_{jl}$ but different for $S_{jl}$. We also recall that, due to the combination of spin-orbit coupling and quantum confinement, the spin of the holes mediating the scattering process and, thus, the electron spin quantization axis are perpendicular to the layers [33]. Since the symmetries of charge and spin excitations are, respectively, $A_1$ and $A_2$, the relevant tensor components for light polarized in the plane of the layers are of the form $C_{x'x'} = C_{y'y'}$ and $S_{x'y'} = -S_{y'x'}$ (all other components vanish). Accordingly, spin oscillations can only be excited by circularly polarized light whereas both linearly and circularly polarized light couple to the charge-density mode. It also follows that the scattered and the incident probe beams must be perpendicular to each other for spin, but they are along the same direction for charge excitations. These selection rules are consistent with the experimental findings.

As mentioned earlier, charge-density excitations are affected by the long range (Hartree) part of the Coulomb interaction whereas spin fluctuations are not screened [32]. The following single-particle analysis provides a simple physical picture of, both, the screening behavior of the two types of collective modes and the associated coherent states created by the laser pulses. Following an impulsive excitation with $\mathbf{q} = 0$, the wavefunction of an electron initially in the state $|\mathbf{k}s,0\rangle$ of the lowest subband becomes, to lowest order in the pump electric field

$$\Psi \approx |\mathbf{k}s,0\rangle e^{-iE_0 t/\hbar} + \sum_{n \neq 0} \alpha_{ns}|\mathbf{k}s,n\rangle e^{-iE_n t/\hbar} \qquad (3)$$

where $\alpha_{ns}$ are constants proportional to the intensity of the pulses ($|\alpha_{ns}| \ll 1$). Hence, the quasi-2DEG density for a given spin polarization varies by

$$\delta\sigma_s(\mathbf{r},t) \approx \sum_{\substack{n \neq 0 \\ k < k_F}} \alpha_{ns}\langle\mathbf{r}|\mathbf{k}s,n\rangle\langle\mathbf{k}s,0|\mathbf{r}\rangle e^{-i(E_n - E_0)t/\hbar} + \text{c.c.} \quad (4)$$

From (2), we have that $\alpha_{n\uparrow} = \pm\alpha_{n\downarrow}$ where the plus (minus) sign is for charge- (spin-) excitations. Thus, the effect of an optical pulse is to create coherent density oscillations for which the spin-up and spin-down components are either in phase (charge-density mode) or $180^{o}$ out of phase (spin-density mode). Because the two contributions add up for charge excitations (i. e., $\delta\sigma_{\uparrow} = \delta\sigma_{\downarrow}$), these plasmon-like modes experience a restoring field whereas spin excitations remain unscreened since the corresponding motion does not change the net density ($\delta\sigma_{\uparrow} + \delta\sigma_{\downarrow} = 0$).

In conclusion, we have shown that ultrafast lasers can be used to generate coherent density oscillations in a 2DEG through stimulated RS, and that charge- and spin-density fluctuations can be independently controlled using the polarization sensitivity of the Raman process.

Acknowledgment is made to the donors of The Petroleum Research Fund, administered by the ACS, for partial support of this research. Work also supported by the NSF Focus Physics Frontier Center.

FIG 1. (a) Sample structure: GaAs (●), Al$_{0.3}$Ga$_{0.7}$As (◎) and Si donors (+) introduced by δ-doping. A smoothing superlattice consisting of 100 periods of 30-Å GaAs and 100-Å Al$_{0.3}$Ga$_{0.7}$As was grown on top of the substrate, which is on the left of the diagram. The two doping layers close to the sample surface help pull the conduction band edge to very near the 2DEG Fermi level at the position of the Si-donors close to the QW. b) Energy level diagram. (c) Photoluminescence spectrum obtained at 7 K with 0.01 W/cm$^2$ of the 4880Å Ar-line. Dashed line denotes the QW gap. The weaker peak at ~ 1.516 eV is not associated with the 2DEG. (d) Wavevector dependence of intersubband excitations for $q << k_F$. (e) Raman spectra recorded in the polarized $z(x',x')\bar{z}$ and depolarized $z(x',y')\bar{z}$ backscattering configurations [26]. The (continuous wave) laser energy is 1.556 eV. For clarity, the spectra have been shifted vertically.

FIG 2. Time-resolved differential reflectivity data. Curves are linear prediction fits [28]. The associated Fourier transform spectra show peaks due to coherent charge and spin-density oscillations. The top and bottom traces were obtained with linearly (parallel to $x'$) and circularly polarized pump pulses. For the probe, the scattered beam polarization is parallel (top) and perpendicular (bottom) to the polarization of the incident beam. Inset: Dependence of the spontaneous CD$_{01}$ Raman cross section on laser energy. Results obtained with a continuous wave Ti-sapphire laser at ~ 1 W/ cm$^2$. Also shown is the energy spectrum of the light pulses used to generate coherent oscillations.

Fig. 3. (a) Differential reflectivity data showing generation of charge-density oscillations at two pump power densities: $U = 2.0 \times 10^{-7}$ J/cm$^2$ (top) and $U = 2.4 \times 10^{-8}$ J/cm$^2$ (bottom), and corresponding Fourier spectra. The peak at 3.97 THz is due to transitions of photoexcited

electrons from the first to the second excited state of the QW. (b) Polarized Raman spectra showing the emergence of the $CD_{12}$ peak at high power densities. The laser energy is 1.610 eV.

FIG. 4. Transient reflectivity changes for double pulse excitation showing destructive (top) and constructive (bottom) interference for $CD_{12}$. Arrows indicate the times at which the two pump pulses were applied. The associated spectra in the insets were obtained by Fourier transforming data for times larger than the one denoted by the dashed line.

FIGURE 1

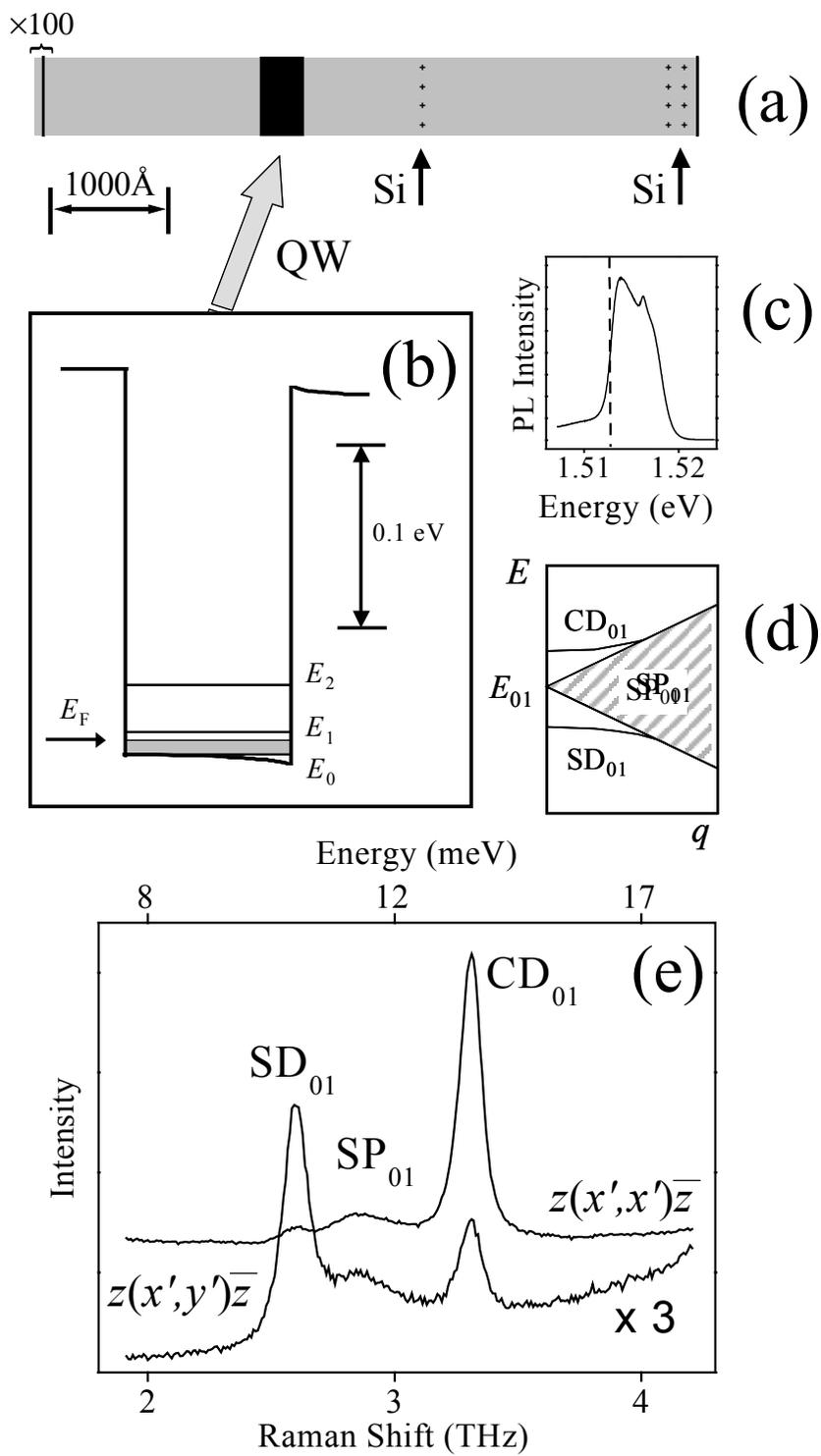

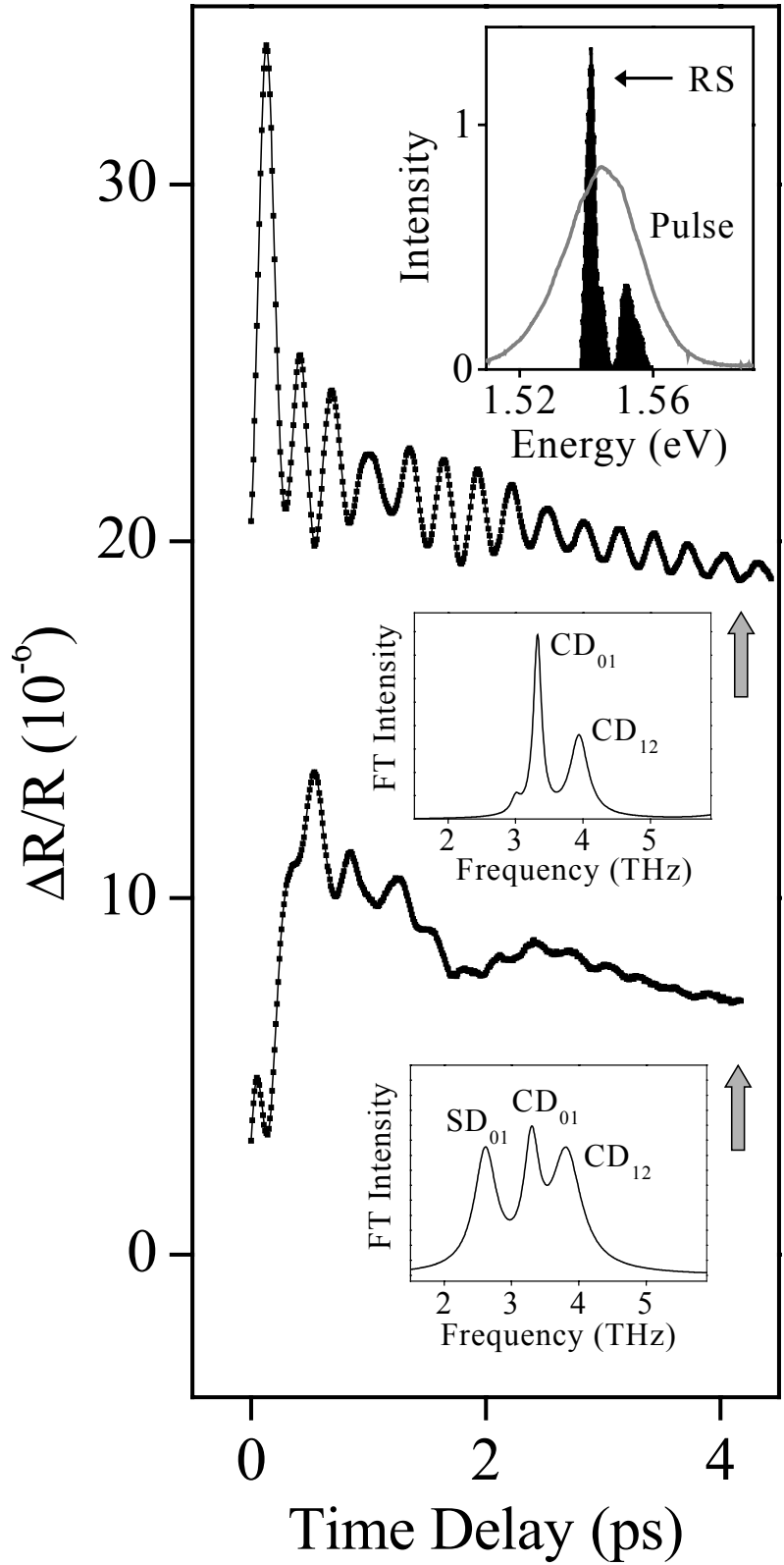

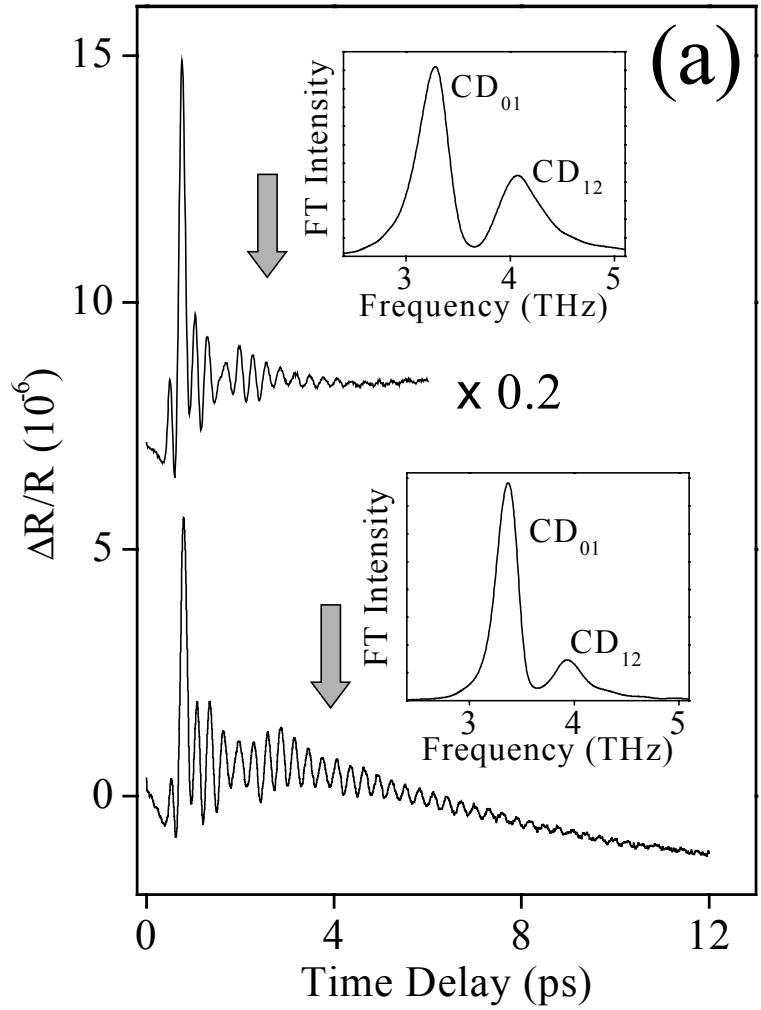

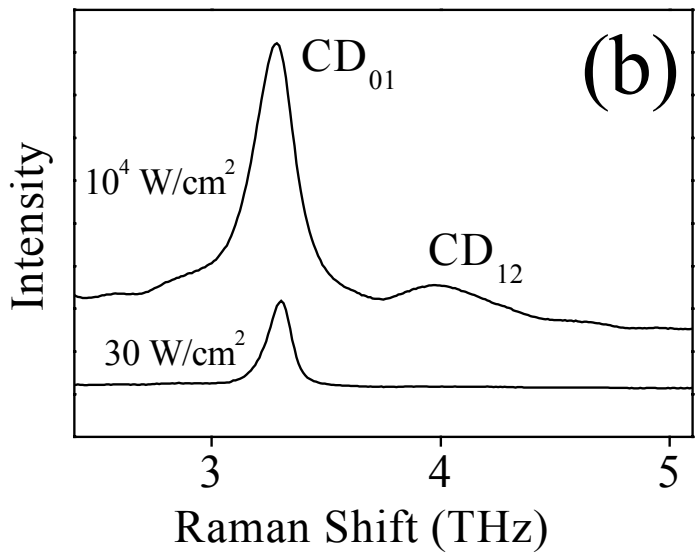

FIGURE 4

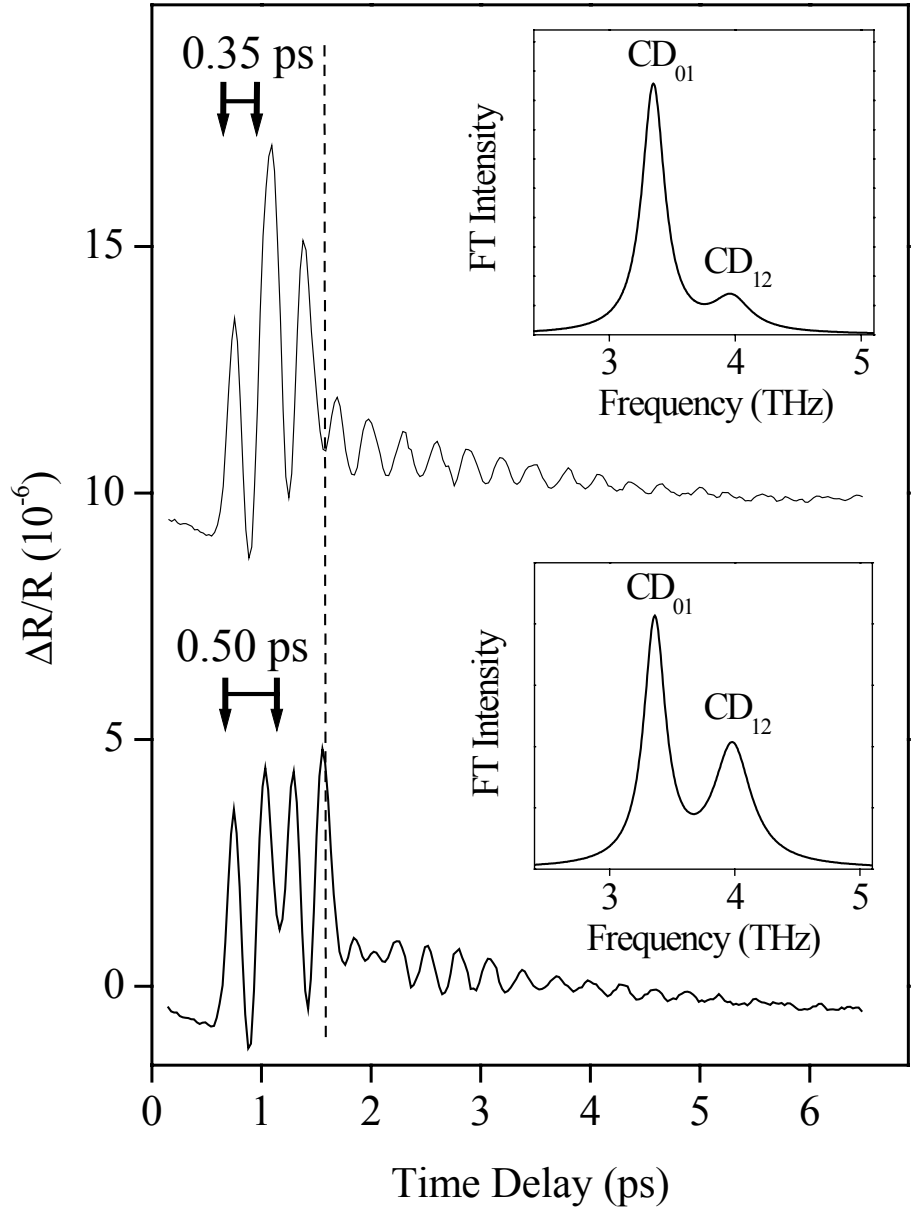